\documentclass[runningheads]{llncs}
\usepackage{graphicx}
\usepackage{url}

\begin{document}
\title{Multi-View Attention for gestational age at birth prediction}
%
%
\author{Mathieu Leclercq\inst{1} \and
Martin Styner\inst{1} \and
Juan Carlos Prieto\inst{1}}
\authorrunning{M. Leclercq et al.}
%
\institute{University of North Carolina, Chapel Hill, NC 27514 USA 
\email{leclercqmathieu14@gmail.com}\\
\url{https://grand-challenge.org/algorithms/multi-view-attention-for-ga-at-birth-regression/}}
%
\maketitle              
\begin{abstract}
We present our method for gestational age at birth prediction for the 
SLCN (surface learning for clinical neuroimaging) challenge. 
Our method is based on a multi-view shape analysis technique that captures 2D renderings 
of a 3D object from different viewpoints. We render the brain features on the surface of the sphere and then the 2D images are analyzed via 
2D CNNs and an attention layer for the regression task. 
The regression task achieves a MAE of $1.637 \pm 1.3$ on the Native space and MAE of $1.38 \pm 1.14$ on the template space. 
The source code for this project is available in our github repository\footnote{\url{https://github.com/MathieuLeclercq/SLCN_challenge_UNC}}. 
\end{abstract}
%
%
%


\section{Data pre-processing and augmentation}


There are no pre-processing steps.
The data augmentation consists of a Dropout layer applied to the inputs and additive Gaussian noise. 

\section{Method description}

\subsection{3D shape analysis}

Learning-based methods for shape analysis use the 3D models to learn descriptors
directly from them. There are mainly 3 types of learning-based methods: multi-view, 
volumetric, multi-layer-perceptrons (MLP). 

Multi-view approaches adapt state-of-the art 2D CNNs to work on 3D shapes. 
The main impediment is the arbitrary structures of 3D models which are usually represented 
by point clouds or triangular meshes, whereas the majority of deep learning algorithms
use the regular grid-like structures found in 2D/3D images\cite{boubolo2021flyby,deleat2021merging}. By rendering 3D objects from 
different view points, features are extracted using 2D CNNs\cite{su2015multi,kanezaki2018rotationnet,ma2018learning}. 
On the other hand, volumetric approaches use
3D voxel grids to represent the shape and apply 3D convolutions
to learn shape features\cite{wu20153d,wang2017cnn,riegler2017octnet}. Finally, other approaches consume the point clouds directly and implement multi-layer-perceptrons and/or transformer architectures, or a generalization of typical CNNs \cite{qi2017pointnet,lian2020deep,li2018pointcnn,wu2019pointconv}.

\begin{figure}[ht]
\centering
    \includegraphics[width=0.49\textwidth]{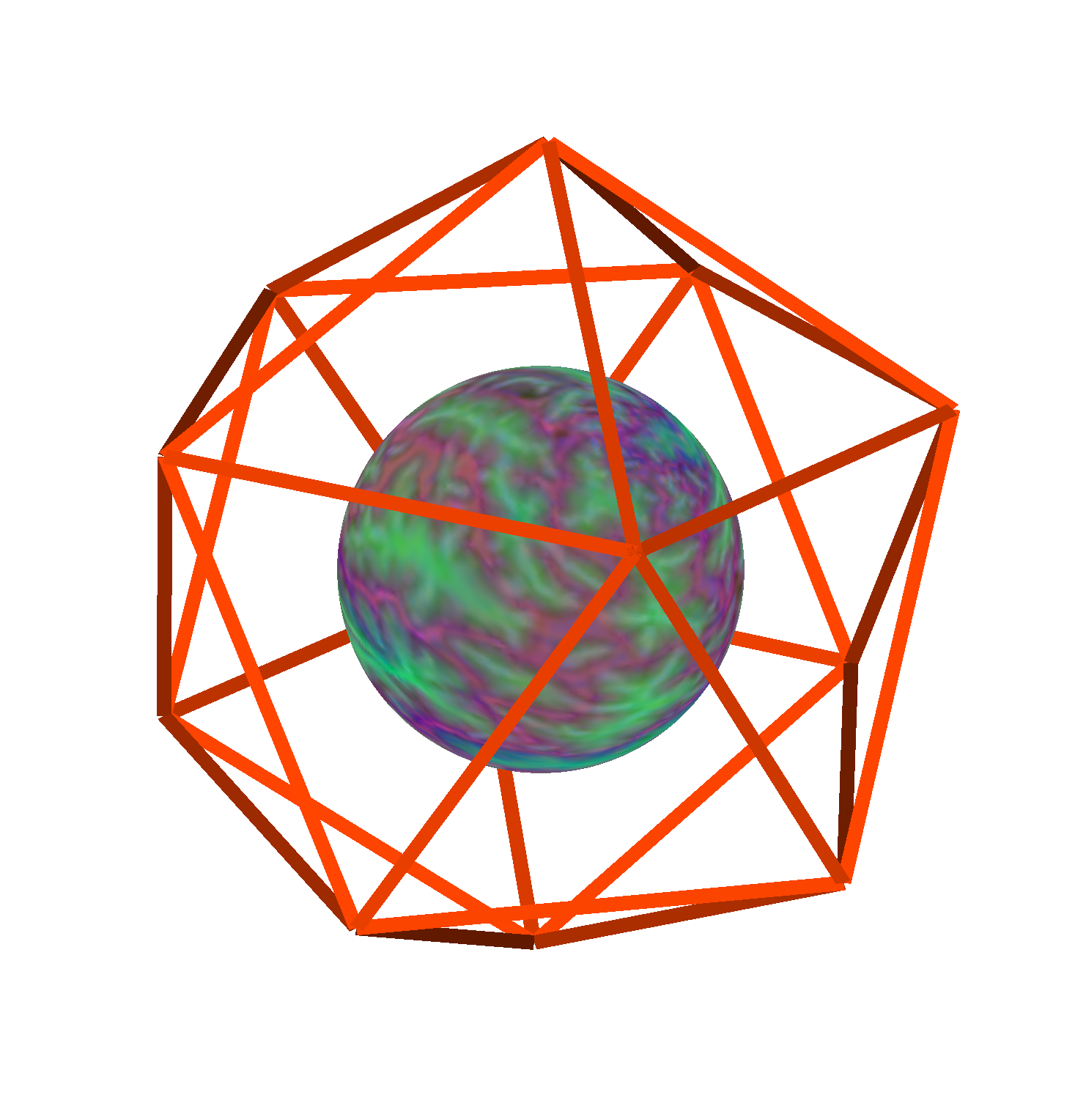}\\
    \includegraphics[width=0.14\textwidth]{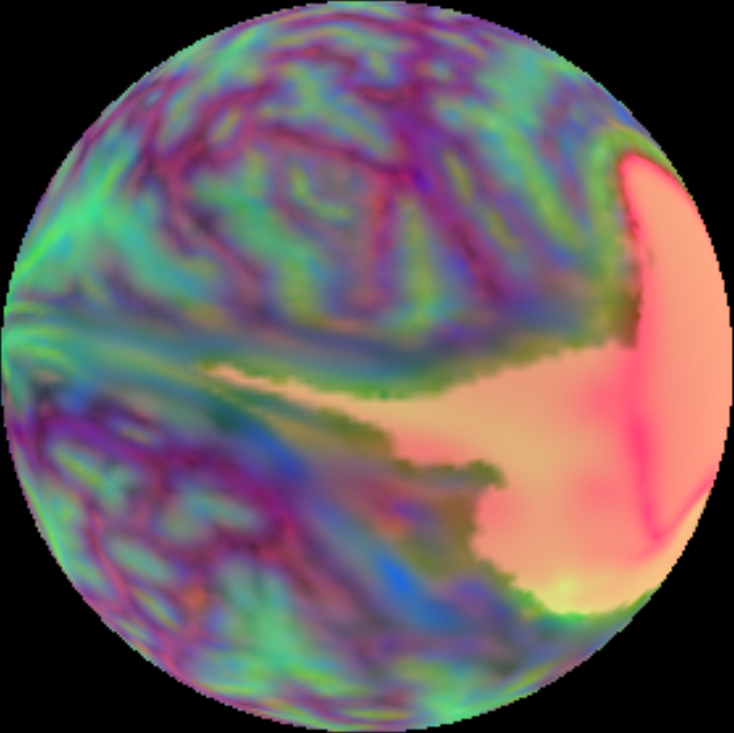}
    \includegraphics[width=0.14\textwidth]{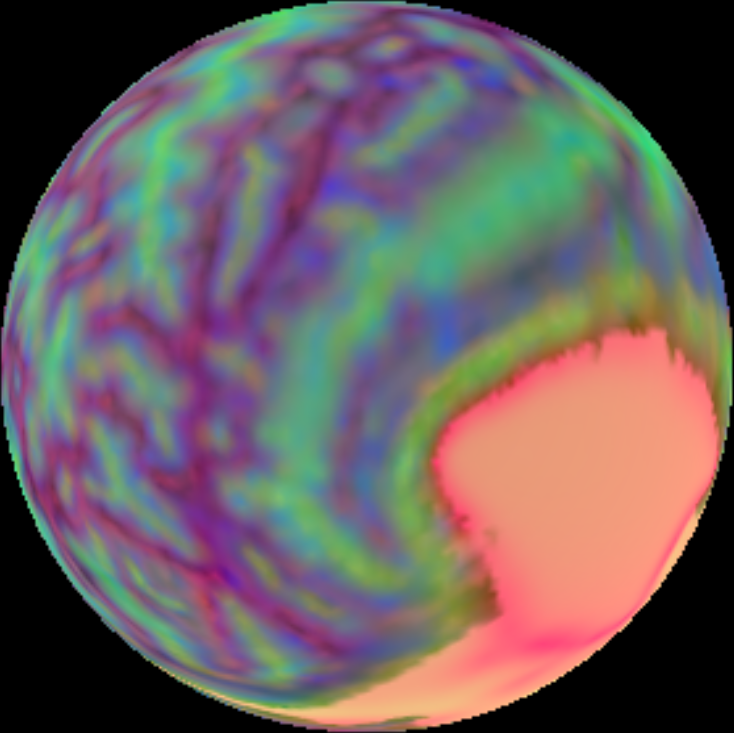}
    \includegraphics[width=0.14\textwidth]{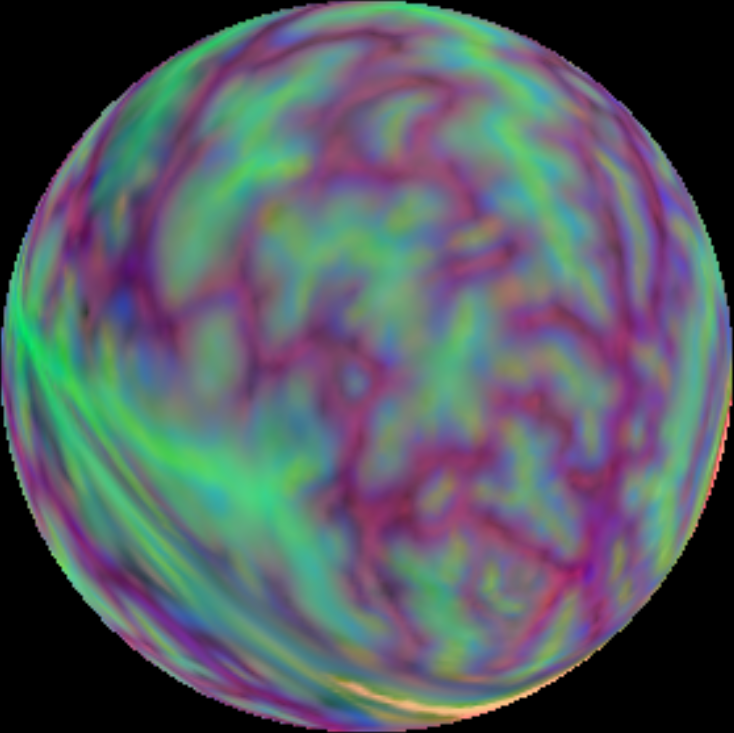}
    \includegraphics[width=0.14\textwidth]{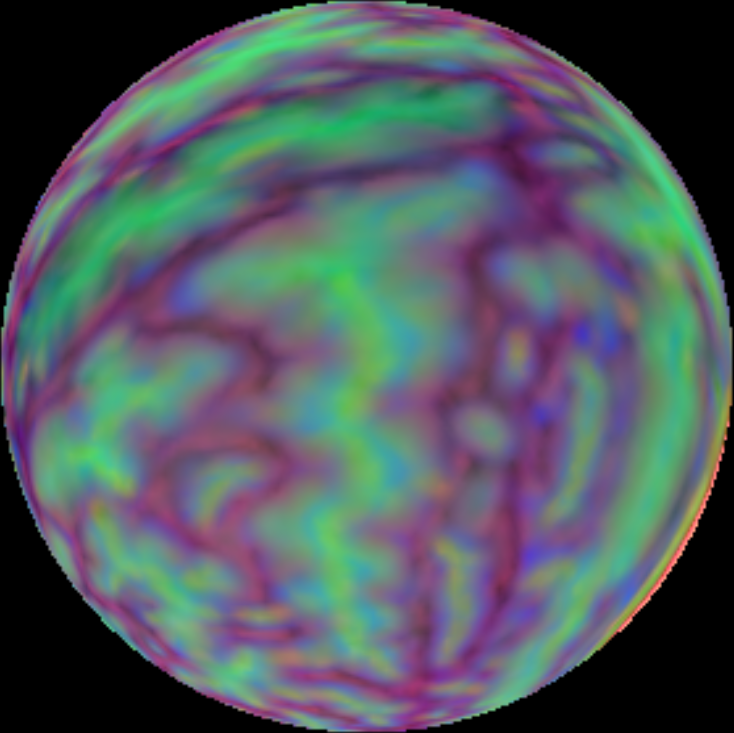}
    \includegraphics[width=0.14\textwidth]{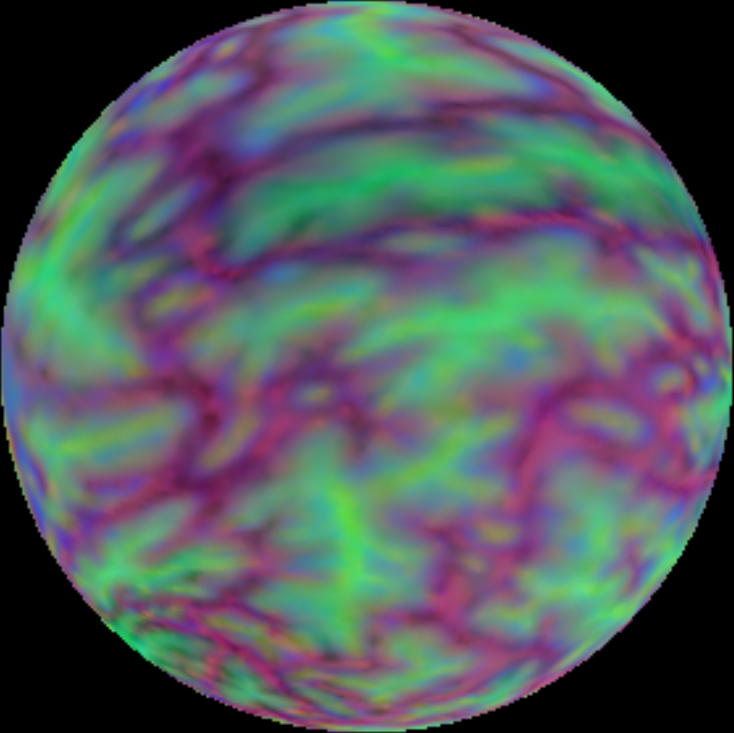}
    \includegraphics[width=0.14\textwidth]{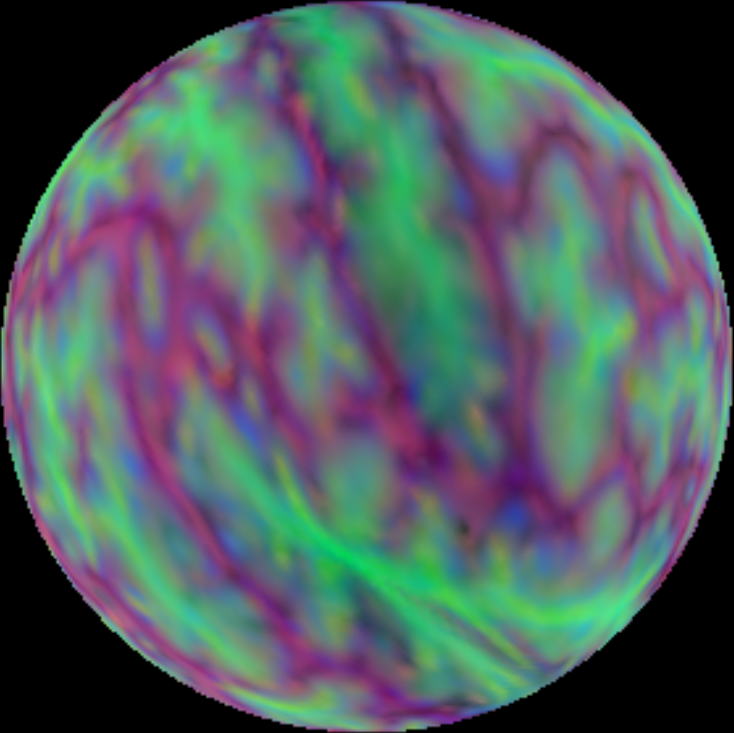}
    \includegraphics[width=0.14\textwidth]{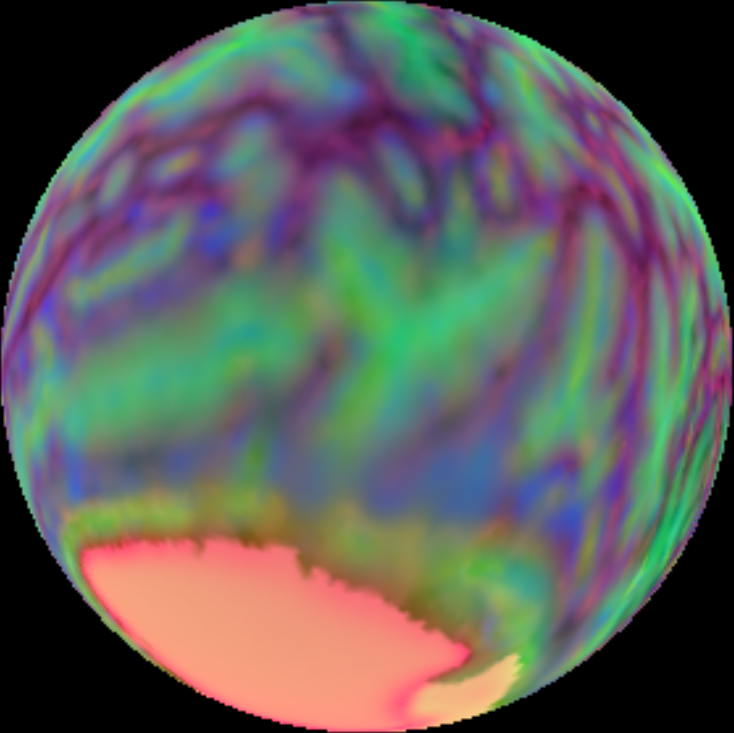}
    \includegraphics[width=0.14\textwidth]{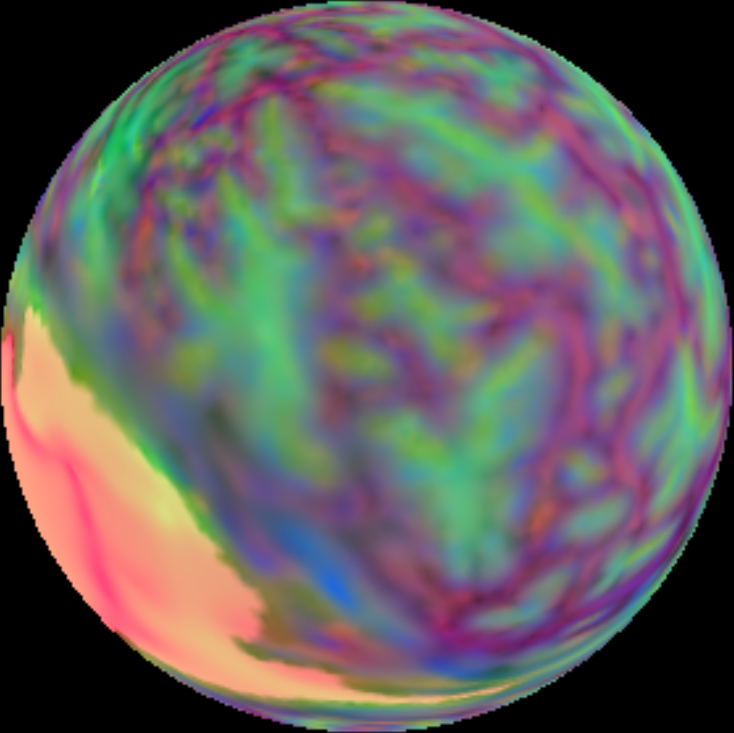}
    \includegraphics[width=0.14\textwidth]{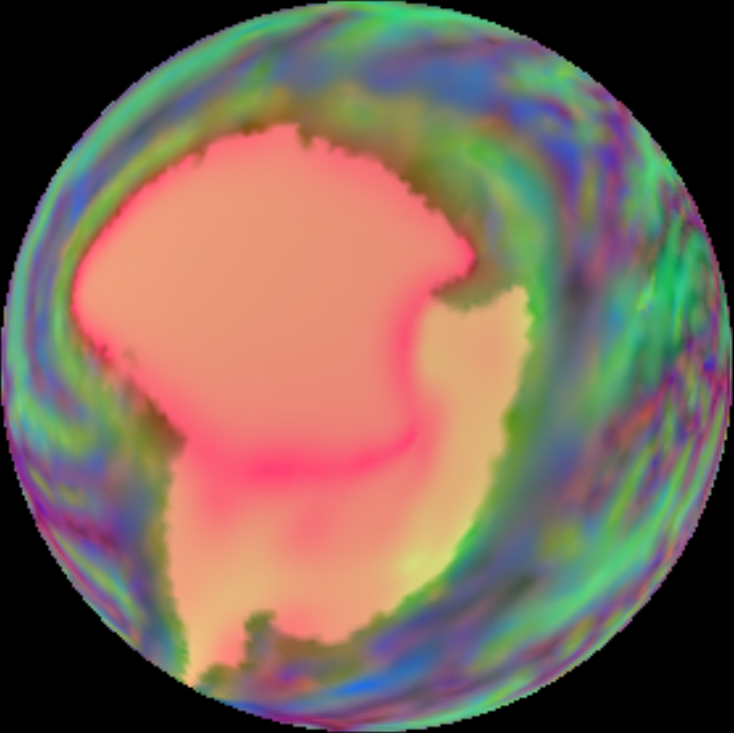}
    \includegraphics[width=0.14\textwidth]{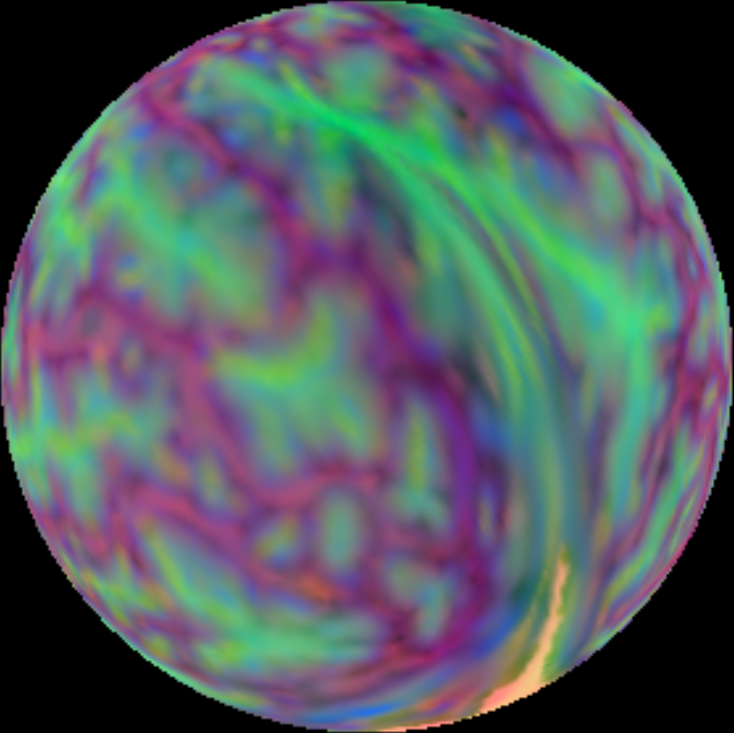}
    \includegraphics[width=0.14\textwidth]{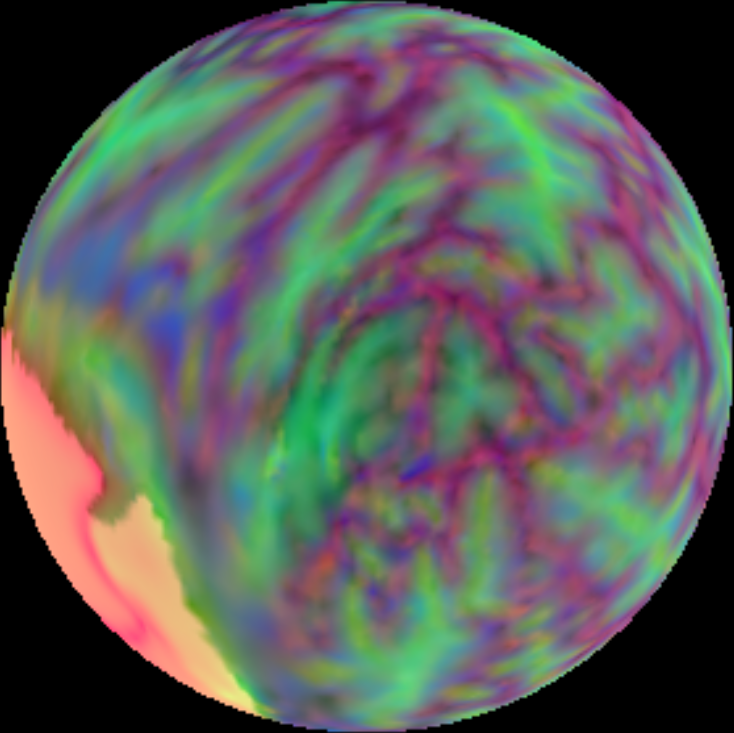}
    \includegraphics[width=0.14\textwidth]{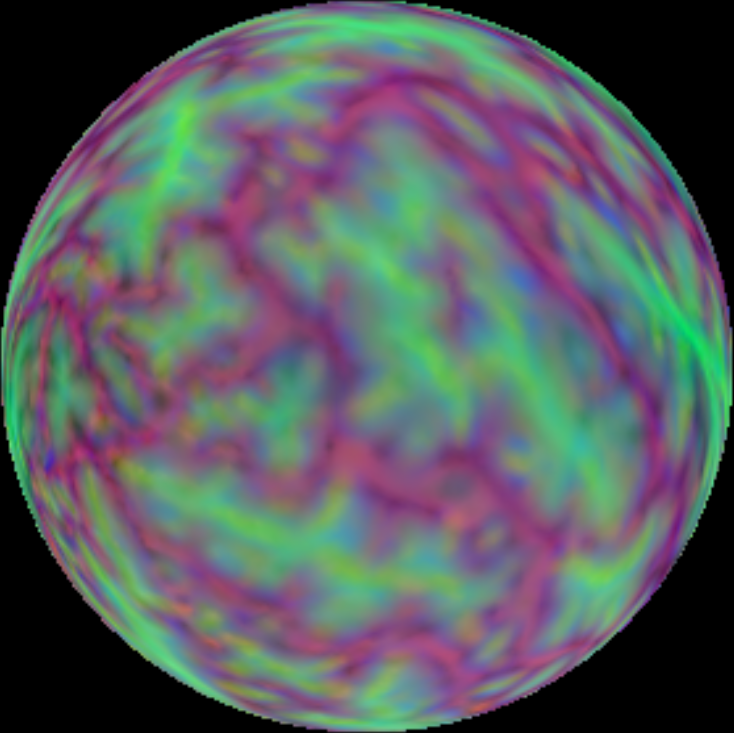}
\caption{Feature rendering on the sphere's surface with the icosahedron subdivision. The rendered 2D images are shown in the bottom rows. These images are used to train our neural network.}
\label{fig:feature_rendering}
\end{figure}

Our method falls in the multi-view category. We render the sphere with the input features 
and capture 2D rendering from viewpoints following the icosahedron subdivision level 0. Figure \ref{fig:feature_rendering} shows the rendering of the sphere and an icosahedron that is used to guide the location of the camera.   

\subsection{Rendering the 2D views}
The Pytorch3D framework allows rendering and training in an end-to-end fashion. 
The rendering engine provides a map that relates pixels 
in the images to faces in the mesh and allows rapid extraction of point data (normals, curvatures, labels, etc.). In this task, we extract the values for the 4 brain features given at each vertex.
The generated images have 4 components and they are fed to the feature extraction network.

We set the resolution of the rendered images to 224px. We use ambient lights so that the rendered images don't have any specular components.

\subsection{Training the neural network}

\begin{figure}[ht]
\centering
    \includegraphics[width=0.75\textwidth]{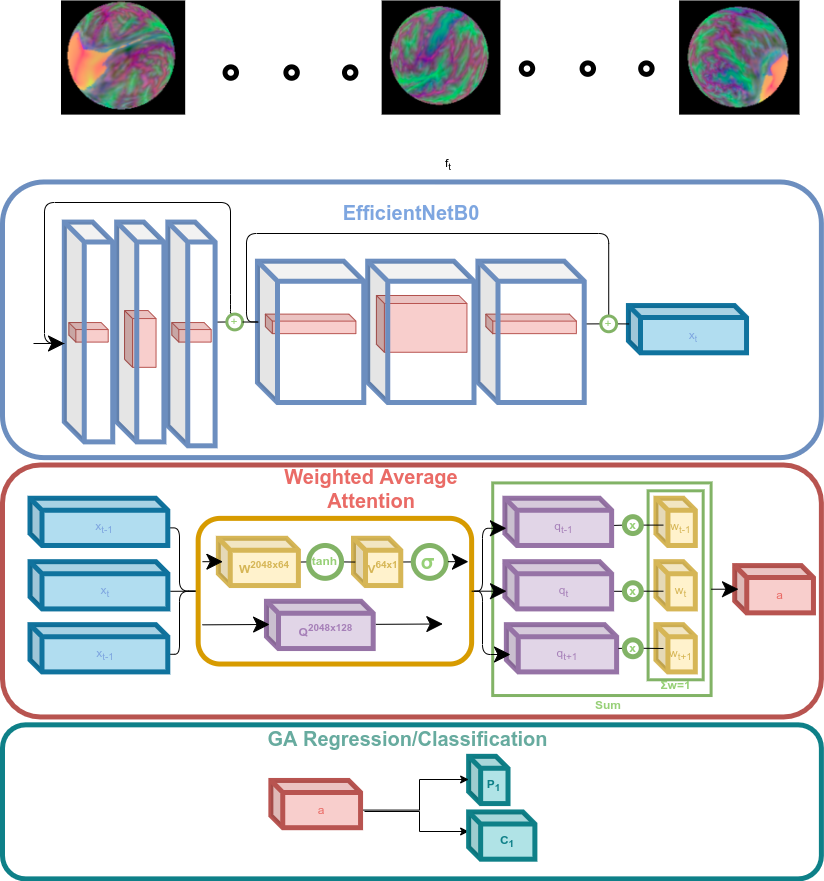}
    \caption{Architecture for gestational age at birth prediction. We use efficientnet B0 as feature extraction and weighted attention layer followed by a linear layer to predict GA.}
    \label{fig:architecture}
\end{figure}
Our model architecture is shown in Figure \ref{fig:architecture}. The model is trained in
an end-to-end fashion using 1 GPU NVIDIA TITAN RTX 24 GB, batch size 18, Adam optimizer, learning rate $1e-4$, Dropout 0.2. We use the early stopping criteria to track the validation loss and save the best performing model. We use data-binning to create classes for the samples in the data set. We use 5 different age bins, \textit{i.e.}, 5 classes ([23 - 27], [27, 32], [32, 36], [36, 40], [40, 44]). This binning step allows us to create weights for the under represented classes as well as adding a new term in our loss function. The loss function is MSE for the regression task plus a weighted classification Cross-Entropy loss. We stop the training using the early stopping criteria after 244 epochs and use the MSE error as validation criteria. 

Our network architecture uses efficient net B0\cite{tan2019efficientnet} to extract image features from the 2D renderings. Then we compute a score for each image and concatenate them to compute a weighted average of the features. The final tensor is then fed to a linear layer that performs the
regression and classification task jointly. 

We use both native and template features to train a single model for the regression task. 

\section{Post processing}

There are no post-processing steps. 

\section{Results}

\begin{table}[]
    \centering
    \begin{tabular}{||c c||} 
         \hline
         \textbf{Space} & MAE $\pm$ STDEV\\ [0.5ex] 
         \hline\hline
         Native & $1.637 \pm 1.250$  \\ 
         \hline
         Template   & $1.380 \pm 1.143$ \\[1ex] 
         \hline
    \end{tabular}
    \caption{MAE and STDEV}
    \label{tab:error}
\end{table}

\begin{figure}[ht]
\centering
    \includegraphics[width=0.8\textwidth]{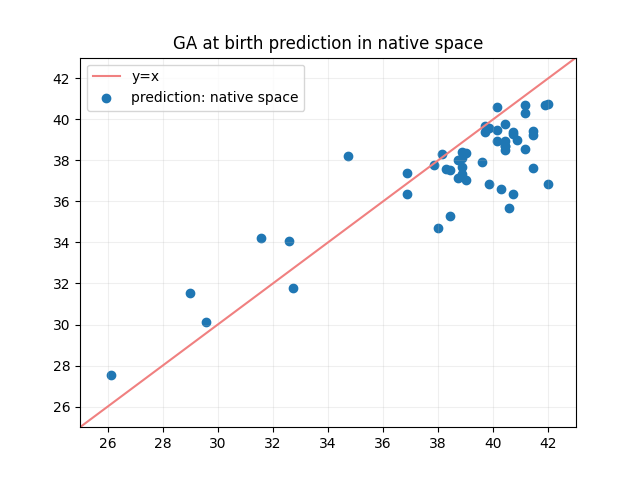}\\
    \includegraphics[width=0.8\textwidth]{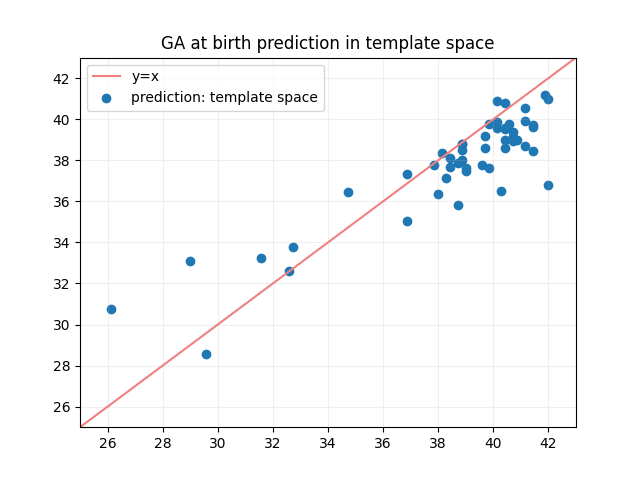}

\caption{Results (in weeks) on the validation set for prediction in native space and template space.}
\label{fig:prediction_results}
\end{figure}

Figure \ref{fig:prediction_results} show that results for native space and template space are very similar because the network was trained using both spaces. The network is slightly biased towards the late GA at birth because there are more sample points in the data set.

\begin{figure}[ht]
\centering
    \includegraphics[width=0.8\textwidth]{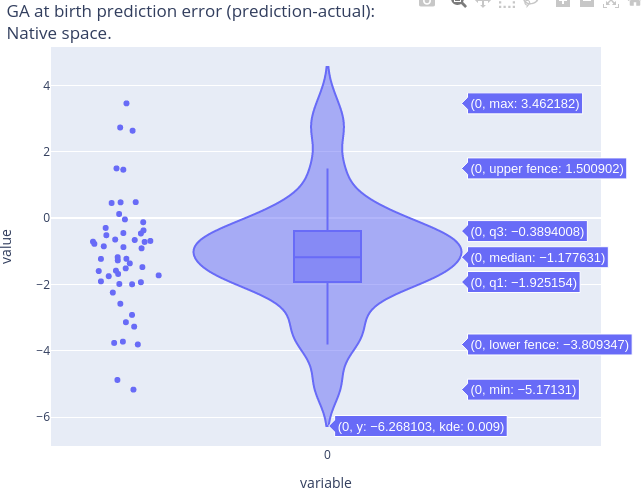}
\caption{Error distribution (in weeks) for prediction in native space. }
\label{fig:violin_native}
\end{figure}

\begin{figure}[ht]
\centering
    \includegraphics[width=0.8\textwidth]{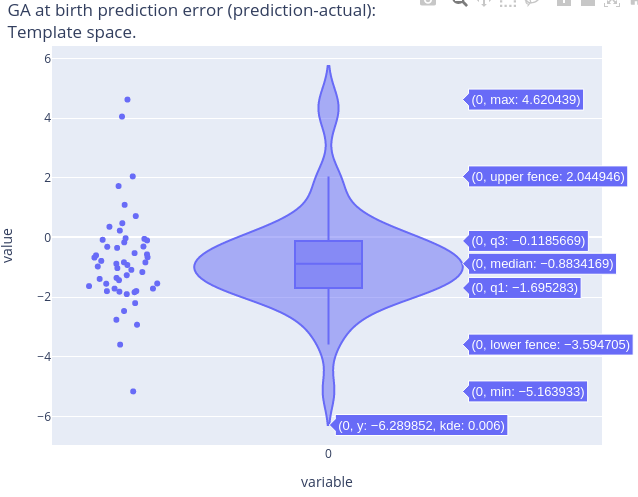}
\caption{Error distribution (in weeks) for prediction in template space.}
\label{fig:violin_template}
\end{figure}

Figures \ref{fig:violin_native} and \ref{fig:violin_template} show that $2/3$ of the validation samples show an absolute error of less than 2 weeks. The model tends to under-predict in both template and native spaces. We consider this is an indication that the model is agnostic to the template/native features. 
Table \ref{tab:error} shows the MAE and STDEV for the prediction task in weeks. We report the results on the validation set. 

%
%
%
\bibliographystyle{splncs04}
\bibliography{surface_learning}

\end{document}